\documentclass[amsmath,amssymb]{revtex4}
\begin{document}
\title{The Klein-Gordon equation with a generalized Hulth\'en potential in $D$-dimensions}
\author{Nasser Saad}
\email{nsaad@upei.ca}
\affiliation{Department of Mathematics and Statistics,
University of Prince Edward Island,
550 University Avenue, Charlottetown,
PEI, Canada C1A 4P3.}


\def\dbox#1{\hbox{\vrule  
                        \vbox{\hrule \vskip #1
                             \hbox{\hskip #1
                                 \vbox{\hsize=#1}%
                              \hskip #1}%
                         \vskip #1 \hrule}%
                      \vrule}}
\def\qed{\hfill \dbox{0.05true in}}  
\def\square{\dbox{0.02true in}} 
\begin{abstract}

\noindent{\bf Abstract:} An approximate solution of the  Klein-Gordon equation for the general Hulth\'en-type potentials in $D$-dimensions within the framework of an approximation to the centrifugal term is obtained. The bound state energy eigenvalues and the normalized eigenfunctions are obtained in terms of hypergeometric polynomials. 
\end{abstract}

\maketitle
\noindent {\bf keyword:}
Hulth\'en potential; Klein-Gordon equation; bound states; Approximate analytic solution, Normalization constant, hypergeometric functions, Incomplete Beta function.\\


\section{Introduction}	
\noindent The search for exact solutions of wave equations, whether non-relativistic or relativistic, has been an important research area since the birth of quantum mechanics. The generalized Hulth\'en potential \cite{hulthen}-\cite{Qiang} is given by 
\begin{equation}\label{eq1}
V(r)=-Z\alpha {e^{-\alpha r}\over 1-qe^{-\alpha r}}
\end{equation}
where $\alpha$ is the screening parameter and $Z$ is a constant which is identified with the atomic number when the potential is used for atomic phenomena. The Hulth\'en potential is one of the important short-range potentials which behaves like a Coulomb potential for small values of $r$ and decreases exponentially for large values of $r$. The Hulth\'en potential  has received extensive study in both relativistic and non-relativistic quantum mechanics \cite{hulthen}-\cite{Qiang}. There is a wealth of literature on the use of the Hulth\'en potential as an approximation of the interaction potential in a number of area in physics such as nuclear and particle physics \cite{sugawara}, atomic  physics \cite{tietz}-\cite{varshni}, solid state physics \cite{Berezin} and chemical physics \cite{Pyykko}, see also \cite{chetouani} and the references therein. Unfortunately, quantum mechanical equations with the Hulth\'en potential can be solved analytically only for states with zero-angular momentum \cite{hulthen}-\cite{Benamira}. Recently, some interesting research papers \cite{chen}-\cite{Qiang} have appeared to study the $l$-state solutions of the relativistic Klein-Gordon equation with Hulth\'en-type potentials. The main idea of their investigation relies on using the approximation of the centrifugal term $1\over r^2$ by means of ${1\over r^2}\approx{\alpha^2 e^{-\alpha r}\over (1-e^{-\alpha r})^2}.$ Their results show that this approximation is in good agreement with the other methods for small $\alpha$ values. The purpose of the present work is twofold: (i) to extend the $l$-state approximate solutions  \cite{chen}-\cite{Qiang} of Klein-Gordon equation with the generalized Hulth\'en potentials to arbitrary dimension; (ii) to compute the normalization constant of the approximate wave functions that seems to have been overlooked by many researchers \cite{hulthen}-\cite{Qiang}.  
\vskip 0.1true in

\section{The Klein-Gordon equation in $D$ dimensions} 
\noindent The $D$-dimensional Klein-Gordon equation for a particle of mass $M$ with radially symmetric Lorentz vector and Lorentz scalar potentials,  $V(r)$ and $S(r)$, $r = \|{\mathbf r}\|,$ is given (in atomic units $\hbar=c=1$) \cite{greiner,alhaidari} by
\begin{equation}\label{eq2}
\{-\Delta_D+[M+S(r)]^2\}\Psi({\bf r})=[E-V(r)]^2\Psi({\bf r}),
\end{equation}
where $E$ denotes the energy and $\Delta_D$ is the $D$-dimensional Laplacian. Transforming to the $D$ dimensional spherical coordinates $(r,\theta_1\dots\theta_{D-1})$, the variables can be separated using
\begin{equation}\label{eq3}
\Psi({\bf r})=R(r)Y_{l_{D-1},\dots,l_1}(\theta_1\dots\theta_{D-1})
\end{equation}
where $R(r)$ is a radial function, and $Y_{l_{D-1},\dots,l_1}(\theta_1\dots\theta_{D-1})$ is a normalized  hyper-spherical harmonic with eigenvalue $l(l+D-2)$, $l=0,1,2,\dots$.  Thus, we obtain the radial equation of Klein-Gordon equation in $D$ dimensions by substituting Eq.(\ref{eq2}) into Eq.(\ref{eq1})
\begin{equation}\label{eq4}
\bigg\{-{d^2\over d r^2}-{{D-1}\over r}{d \over dr}+{l(l+D-2)\over r^2}+[M+S(r)]^2-[E-V(r)]^2\bigg\}R({\bf r})=0.
\end{equation}
 Writing $R$ as $R(r)=r^{-(D-1)/2}u(r)$ gives
\begin{eqnarray}\label{eq5}
\bigg\{-{d^2\over dr^2}&+&{(k-1)(k-3)\over 4r^2}+\big[[M+S(r)]^2-[E-V(r)]^2\big]\bigg\}u(r)=0.
\end{eqnarray}
where $k=D+2l$ and $u(r)$ is the reduced radial wave function satisfying $u(0) = 0.$ 

\section{Bound states for generalized Hulth\'en Potential} 
\noindent We consider the vector and scalar Hulth\'en potential defined as,
\begin{equation}\label{eq6}
V(r)=-{V_0e^{-\alpha r}\over 1-qe^{-\alpha r}},\quad S(r)=-{S_0e^{-\alpha r}\over 1-qe^{-\alpha r}},
\end{equation}
where $V_0$ and $S_0$ are the depth of the vector and scalar Hulth\'en potential respectively and $\alpha$ is the screening parameter and $q\neq 0$ is the deformation parameter. Substituting (\ref{eq6}) in the Klein Gordon equation (\ref{eq5}), we obtain
\begin{equation}\label{eq7}
\bigg\{-{d^2\over dr^2}+{(k-1)(k-3)\over 4r^2}-{(2M S_0+2E V_0)e^{-\alpha r}\over 1-qe^{-\alpha r}}+
{(S_0^2-V_0^2)e^{-2\alpha r}\over (1-qe^{-\alpha r})^2}
\bigg\}u(r)=(E^2-M^2)u(r).
\end{equation}
In order to obtain analytic solutions of this equation, we have to use an approximation \cite{chen}-\cite{Qiang} for the centrifugal term similar to that used for the non-relativistic cases. We, thus, follow \cite{chen}-\cite{Qiang}, \cite{ikhdair}-\cite{bayrak} and use
${1\over r^2}\approx{\alpha^2 e^{-\alpha r}\over (1-qe^{-\alpha r})^2}$ for the centrifugal term. This approximation is valid for $q=1$, however, we follow the model used by Qiang \emph{et al} \cite{Qiang}. This  allow us to write Eq.(\ref{eq7}) as
\begin{equation}\label{eq8}
\bigg\{-{d^2\over dr^2}-{(2M S_0+2E V_0)e^{-\alpha r}\over 1-qe^{-\alpha r}}+
{{\alpha^2\over 4}{(k-1)(k-3)}e^{-\alpha r}+(S_0^2-V_0^2)e^{-2\alpha r}\over (1-qe^{-\alpha r})^2}
\bigg\}u(r)=(E^2-M^2)u(r).
\end{equation}
Eq.(\ref{eq8}) can be further simplified using a new variable $z=qe^{-\alpha r}$ ($r\in [0,\infty), z\in [q,0)$),
\begin{equation}\label{eq9}
\bigg\{{d^2\over dr^2}+{1\over z}{d\over dz}-{\epsilon^2\over z^2}+{\beta_1\over z(1-z)}-{{1\over 4}(k-1)(k-3)\over qz(1-z)^2}-{\beta_2\over (1-z)^2}
\bigg\}u(z)=0,
\end{equation}
where we used the dimensionless parameters given by 
\begin{equation}\label{eq10}
\epsilon={\sqrt{M^2-E^2}\over \alpha},\quad\quad
\beta_1={2(MS_0+EV_0)\over \alpha^2 q},\quad\quad
 \beta_2={S_0^2-V_0^2\over \alpha^2 q^2}.
\end{equation}
We now look for a solution of (\ref{eq9}) in the form
\begin{equation}\label{eq11}
u(z)=z^\epsilon(1-z)^\delta f(z).
\end{equation}
In this case, Eq.(\ref{eq9}) reads 
\begin{eqnarray}
f''(z)
&+&\bigg({1+2\epsilon\over z}-{2\delta\over 1-z}\bigg)f'(z)\nonumber\\
&+&\bigg(
{{(\beta_1-(2\epsilon+1)\delta-\delta^2+\delta-\beta_2)(1-z)+\delta^2-\delta-\beta_2-{1\over 4q}(k-1)(k-3)}\over z(1-z)^2}\bigg)f(z)=0\nonumber\\
\label{eq12}
\end{eqnarray}
We may now choose $\delta$ such that
\begin{equation}\label{eq13}
\delta^2-\delta-\beta_2-{1\over 4q}(k-1)(k-3)=0
\end{equation}
which yields
\begin{equation}\label{eq14}
\delta=\delta_{\pm}={1\over 2}\pm {1\over 2q} \sqrt{q^2(1+4\beta_2)+q(k-1)(k-3)}
\end{equation}
where $\delta=\delta_+$ for $q>0$ and $\delta=\delta_-$ for $q<0$.
Thus Eq.(\ref{eq12}) reduce to
\begin{equation}\label{eq15}
f''(z)
=\bigg({2\delta\over 1-z}-{1+2\epsilon\over z}\bigg)f'(z)
+\bigg(
{4q(\delta(1+2\epsilon)-\beta_1)+(k-1)(k-3)\over 4qz(1-z)}\bigg)f(z)
\end{equation}
This equation is a special case of a more general differential equation discussed in \cite{Hakan}, namely 
\begin{equation}\label{eq16}
f''(z)
=\bigg({2az^{N+1}\over 1-bz^{N+2}}-{2(m+1)\over z}\bigg)f'(z)
-{wz^N\over (1-bz^{N+2})}f(z), \quad N=-1,0,1,\dots
\end{equation}
which has exact solutions, for $n=0,1,2,\dots,$ given by
\begin{equation}\label{eq17}
f_n(z)=(-1)^nC_n(N+2)^n\left({2m+N+3\over N+2}\right)_n{}_2F_1(-n,{(2m+1)b+2a\over (N+2)b}+n; {2m+N+3\over N+2}; b z^{N+2})
\end{equation}
if
\begin{equation}\label{eq18}
w_n(N)=n(N+2)((n(N+2)+{(2m+1))b+2a}).
\end{equation}
Here $C_n$ is the normalization constant and ${}_2F_1(a,b;c;n)$ is a special case \cite{andr} of the generalized hypergeometric function 
\begin{equation}\label{eq19}
{}_pF_q(a_1,...,a_p;c_1,...,c_q;z) = \sum_{n=0}^\infty
\frac{(a_1)_n\cdot\cdot\cdot(a_p)_n}{(c_1)_n\cdot\cdot\cdot(c_q)_n}\frac{z^n}{n!}\,
\end{equation}
where the Pochhammer symbol $(a)_n$ is defined by $(a)_n=\Gamma(a+n)/\Gamma(a)$.
By putting $N=-1$ and $b=1$ in (\ref{eq16}),  using (\ref{eq18}), with $a=\delta$ and $w_n(-1)=n(n+2\epsilon+2\delta)$, we obtain the energy spectrum of (\ref{eq9}) as
\begin{eqnarray}\label{eq20}
\epsilon_n^{(k)}&=&{q[\beta_1-n^2-(1+2n)\delta]-{1\over 4}(k-1)(k-3)\over 2q(n+\delta)}\nonumber\\
&=&{q[\beta_1-(n+\delta)^2]+q[\delta^2-\delta-{1\over 4q}(k-1)(k-3)]\over 2q(n+\delta)}\nonumber\\
&=&{\beta_1+\beta_2\over 2(n+\delta)}-{1\over 2}(n+\delta)
\end{eqnarray}
where we have used (\ref{eq13}). Furthermore, the exact solutions of (\ref{eq12}), using (\ref{eq17}), now reads
\begin{equation}\label{eq21}
f_n(z)
=(-1)^n~C_n~(1+2\epsilon_n^{(k)})_n~{}_2F_1(-n,2\epsilon_n^{(k)}+2\delta+n;1+2\epsilon_n^{(k)},z)
\end{equation}
Therefore, we can write the total radial wave function (\ref{eq11}) as follows
\begin{eqnarray}\label{eq22}
u_n(z)
&=&C_n~z^{\epsilon_n^{(k)}}(1-z)^\delta~{}_2F_1(-n,2\epsilon_n^{(k)}+2\delta+n;1+2\epsilon_n^{(k)},z)\nonumber\\
&=&C_n~z^{\epsilon_n^{(k)}}(1-z)^\delta~P_n^{(2\epsilon_n^{(k)},2\delta-1)}(1-2z)
\end{eqnarray}
where we used the definition of Jacobi polynomials \cite{andrew} given by
\begin{equation}\label{eq23}
P_n^{(\alpha,\beta)}(z)={\Gamma(n+1+\alpha)\over n!\Gamma(1+\alpha)}{}_2F_1(-n,\alpha+\beta+1+n;1+\alpha,{1-z\over 2}).
\end{equation}
\section{Normalization Constant}
In this section, we compute the normalization constant $C_{n}$ appear in (\ref{eq22}). As far as we aware, the normalized energy eigenfunctions have not been explicitly worked out in the literature for this case. It is straightforward to note that the normalization constant $C_{n}$ can be computed using $\int_0^\infty |R^2(r)|r^{(D-1)}dr=\int_0^\infty |u(r)|^2dr=-\int_q^0|u(z)^2|{dz\over\alpha z}=1$ because of the substitution $z=qe^{-\alpha r}$. Thus, we have by mean of (\ref{eq22}) that the normalization constant $C_n$ is given by
\begin{equation}\label{eq24}
C_n^2\int\limits_0^q z^{2\epsilon_n^{(k)}-1} (1-z)^{2\delta}\left[{}_2F_1(-n,2\epsilon_n^{(k)}+2\delta+n;1+2\epsilon_n^{(k)},z)\right]^2dz=\alpha.
\end{equation}
Using the series representation (\ref{eq19}) of the hypergeometric function ${}_2F_1$, being a polynomial of degree $n$ in $z$,  we have
\begin{equation}\label{eq25}
C_n^2\sum\limits_{i=0}^n\sum\limits_{j=0}^n
{(-n)_i(2\epsilon_n^{(k)}+2\delta+n)_i\over
(1+2\epsilon_n^{(k)})_i i!}
{(-n)_j(2\epsilon_n^{(k)}+2\delta+n)_j\over
(1+2\epsilon_n^{(k)})_j j!}\int\limits_0^q z^{2\epsilon_n^{(k)}+i+j-1} (1-z)^{2\delta}dz=\alpha.
\end{equation}
The definite integral in (\ref{eq25}) is just the integral representation of Incomplete Beta function \cite{temme}, 
\begin{equation}\label{eq26}
B_q(x,y)=\int_0^q t^{x-1}(1-t)^{y-1}dt,\quad \Re(x)>0,\Re(y)>0.
\end{equation}
Therefore, Eq.(\ref{eq25}) now reads
\begin{equation}\label{eq27}
C_n^2\sum\limits_{i=0}^n\sum\limits_{j=0}^n
{(-n)_i(2\epsilon_n^{(k)}+2\delta+n)_i\over
(1+2\epsilon_n^{(k)})_i i!}
{(-n)_j(2\epsilon_n^{(k)}+2\delta+n)_j\over
(1+2\epsilon_n^{(k)})_j j!}B_q(2\epsilon_n^{(k)}+i+j,2\delta+1)=\alpha
\end{equation}
On integrating by parts of (\ref{eq26}), we can find that the Incomplete Beta function satisfies the recurrence relation
\begin{equation}\label{eq28}
B_q(x+1,y)={x\over x+y}B_q(x,y)-{q^x(1-q)^y\over x+y}.
\end{equation}
which in turn can be written in terms of the normalized version of the Incomplete Beta function $I_q(x,y)=B_q(x,y)/B(x,y)$ as
\begin{eqnarray}\label{eq29}
I_q(x,y)&=&I_q(x-1,y)-{q^{x-1}(1-q)^y\over (x-1)B(x-1,y)}\nonumber\\
&=&I_q(x-2,y)-{q^{x-2}(1-q)^y\over (x-2)B(x-2,y)}-{q^{x-1}(1-q)^y\over (x-1)B(x-1,y)}\nonumber\\
&=&I_q(x-3,y)-{q^{x-3}(1-q)^y\over (x-3)B(x-3,y)}-{q^{x-2}(1-q)^y\over (x-2)B(x-2,y)}-{q^{x-1}(1-q)^y\over (x-1)B(x-1,y)}\nonumber\\
&=&\dots\nonumber\\
&=&I_q(x-m,y)-q^{x}(1-q)^y\sum\limits_{k=1}^m{q^{-k}\over (x-k)B(x-k,y)}, \quad m=1,2,\dots
\end{eqnarray}
Thus
\begin{eqnarray}
I_q(2\epsilon_n^{(k)}+i+j,2\delta+1)
&=&I_q(2\epsilon_n^{(k)},2\delta+1)-q^{2\epsilon_n^{(k)}+i+j}(1-q)^{2\delta+1}\sum\limits_{k=1}^{i+j}{q^{-k}\over (2\epsilon_n^{(k)}+i+j-k)B(2\epsilon_n^{(k)}+i+j-k,2\delta+1)}\nonumber\\
\label{eq30}
\end{eqnarray}
which allow us to compute the Incomplete Beta function in (\ref{eq27}) as $B_q(2\epsilon_n^{(k)}+i+j,2\delta+1)= B(2\epsilon_n^{(k)}+i+j,2\delta+1)I_q(2\epsilon_n^{(k)}+i+j,2\delta+1)$ for $i,j=0,1,2,\dots$. Note, in the case of $i=j=0$, the sum in (\ref{eq30}) is  equal to zero.\\

Although the discussion above for computing the normalization constant assumed that $q\in (0,1)$, the computation for arbitrary $q\neq 0$ can be performed using the analytic expressions \cite{temme}:
\begin{eqnarray}\label{eq31}
B_q(x,y)&=&{q^x(1-q)^{y-1}\over x}\sum_{k=0}^\infty {(1-y)_k\over (1+x)_k}\bigg({q\over q-1}\bigg)^k\nonumber\\
&=&{q^x(1-q)^{y-1}\over x}{}_2F_1(1,1-y;1+x;{q\over q-1})
\end{eqnarray}
for $q\in (-\infty,0)\cup (0,{1\over 2})$, 
and 
\begin{eqnarray}\label{eq32}
B_q(x,y)&=&B(x,y)-{q^{x-1}(1-q)^{y}\over y}\sum_{k=0}^\infty {(1-x)_k\over (1+y)_k}\bigg({q-1\over q}\bigg)^k\nonumber\\
&=&B(x,y)-{q^{x-1}(1-q)^{y}\over y}{}_2F_1(1,1-x;1+y;{q-1\over q})
\end{eqnarray}
for $q>{1\over 2}$. In the case of $q=1$, i.e. the Hulth\'en potential, Eq.(\ref{eq32}) yields $B_1(x,y)=B(x,y)$, consequently, Eq.(\ref{eq25}) becomes
\begin{equation}\label{eq33}
C_{n}^2\sum\limits_{i=0}^n\sum\limits_{j=0}^n
{(-n)_i(2\epsilon_n^{(k)}+2\delta+n)_i\over
(1+2\epsilon_n^{(k)})_i i!}
{(-n)_j(2\epsilon_n^{(k)}+2\delta+n)_j\over
(1+2\epsilon_n^{(k)})_j j!} B(2\epsilon_n^{(k)}+i+j,2\delta+1)=\alpha
\end{equation}
Using the definition of Beta function \cite{temme} in terms of Gamma function $B(x,y)={\Gamma(x)\Gamma(y)\over \Gamma(x+y)}$, we write (\ref{eq33}) as 
\begin{equation}\label{eq34}
C_{n}^2\sum\limits_{i=0}^n\sum\limits_{j=0}^n
{(-n)_i(2\epsilon_n^{(k)}+2\delta+n)_i\over
(1+2\epsilon_n^{(k)})_i i!}
{(-n)_j(2\epsilon_n^{(k)}+2\delta+n)_j\over
(1+2\epsilon_n^{(k)})_j j!}{\Gamma(2\epsilon_n^{(k)}+i+j)\Gamma(2\delta+1)
\over\Gamma(2\epsilon_n^{(k)}+i+j+2\delta+1)}=\alpha
\end{equation}
By means of the definition of Pochhammer symbols $(a)_n={\Gamma(a+n)\over \Gamma(a)}$, we have
\begin{equation}\label{eq35}
C_{n}^2{\Gamma(2\epsilon_n^{(k)})\Gamma(2\delta+1) \over \Gamma(2\epsilon_n^{(k)}+2\delta+1)}
\sum\limits_{i=0}^n
{(-n)_i(2\epsilon_n^{(k)}+2\delta+n)_i(2\epsilon_n^{(k)})_i\over
(1+2\epsilon_n^{(k)})_i(2\epsilon_n^{(k)}+2\delta+1)_ii!}\sum\limits_{j=0}^n
{(-n)_j(2\epsilon_n^{(k)}+2\delta+n)_j(2\epsilon_n^{(k)}+i)_j\over
(1+2\epsilon_n^{(k)})_j (2\epsilon_n^{(k)}+i+2\delta+1)_jj!} 
=\alpha
\end{equation}
Thus, by using the series representation of the hypergeometric series ${}_3F_2$, again Eq.(\ref{eq19}), Eq.(\ref{eq35}) then reduce to
\begin{equation}\label{eq36}
C_{n}^2\sum\limits_{i=0}^n
{(-n)_i(2\epsilon_n^{(k)}+2\delta+n)_i(2\epsilon_n^{(k)})_i\over(2\epsilon_n^{(k)}+2\delta+1)_ii!}~
{}_3F_2\left(\begin{array}{lll}
-n,& 2\epsilon_n^{(k)}+2\delta+n,&2\epsilon_n^{(k)}+i \\
1+2\epsilon_n^{(k)},&2\epsilon_n^{(k)}+i+2\delta+1 \\
\end{array};1
\right)={\alpha\over B(2\epsilon_n^{(k)},2\delta+1) }
\end{equation}
which can be used to compute the normalization constant for $n=0,1,2,\dots$. In particular, for the ground-state $n=0$, we have
\begin{equation}\label{eq37}
C_{0}=\sqrt{{\alpha\over B(2\epsilon_0^{(k)},2\delta+1)}}.
\end{equation}

\section{Conclusion}
\noindent In this work, we have extended the approximate analytic solutions of Klein-Gordon equation with vector and scalar generalized Hulth\'en potential to arbitrary dimension $D$. The analytical energy equation and the normalized radial wave functions expressed in terms of hypergeometric polynomials are given. When $D=3$, our results  normalize  the approximate analytic solution for bound states obtained in \cite{chen} and \cite{Qiang} for Klein-Gordon equation with generalized Hulth\'en potentials for nonzero angular momentum.

\section*{Acknowledgments}
\noindent Partial financial support of this work under Grant No. GP249507 from the 
Natural Sciences and Engineering Research Council of Canada is gratefully 
acknowledged.


\begin{thebibliography}{9}
\bibitem{hulthen} L. Hulth\'en, {\it \"Uber die Eigenl\"osungen der Schr\"odingergleichung des Deuterons
\/}, Arkiv. Mat. Astr. Fysik. 28A(5), (1942) 1-12.

\bibitem{sugawara} L. Hulth\'en,   M. Sugawara,   S. Fl\"ugge (ed.), {\it Handbuch der Physik}, Springer  (1957).
\bibitem{tietz} T. Tietz, {\it Negative Hydrogen Ion}, J. Chem. Phys. 35 (1961) 1917-1918.

\bibitem{varshni} C. S. Lam and Y. P. Varshni,{\it Energies of s Eigenstates in a Static Screened Coulomb Potential,} Phys. Rev. A, 4 (1971) 1875-1881.

 \bibitem{Berezin} A. A. Berezin,  Phys. Status. Solidi (b) , 50  (1972) 71.

 \bibitem{Pyykko} P. Pyykko,   J. Jokisaari,{\it Spectral density analysis of nuclear spin-spin coupling: I. A Hulth\'en potential LCAO model for $J_{X-H}$ in hydrides $XH_4$},  Chem. Phys. 10  (1975)  pp. 293 - 301.

\bibitem{chetouani} L. Chetouani, L. Guechi, A. Lecheheb, T. F. Hammann, and A. Messouber, {\it Path integral for Klein-Gordon particle in vector plus scalar Hulth\'en-type potential}, Physica A 234 (1996) 529-544.
\bibitem{hall} Richard L. Hall, {\it The Yakawa and Hulth\'en potentials in Quantum mechanics}, J. Phys. A: Math. Gen. 25 (1992) 1373-1382.

\bibitem{flugge} S. Fl\"ugge, {\it Practical Quantum Mechanics}, vol. 1, Springer, Berlin (1994). Problem 68, p. 175.

\bibitem{znojil} M. Znojil, {\it Exact solution of the Schr\"odinger and Klein-Gordon equations for generalised Hulth\'en potentials}, J. Phys. A: Math. Gen. 14 (1981) 383-394. 

\bibitem{dom} F. Dom\'inguez-Adame, {\it Bound states of the Klein-Gordon equation with vector and scalar Hulth\'en-type potentials}, Phys. Lett. A 136 (1989) 175-177. 

\bibitem{ypvarshni} Y. P. Varshni, {\it Eigenenergies and oscillator strengths for the Hulth\'en potential},  Phys. Rev. A 41 (1990) 4682.

\bibitem{mehmet} M. Simsek and H. Egrifes, {\it The Klein-Gordon equation of generalized Hulthen potential in complex quantum mechanics,} J. Phys. A, Math. Gen. 37 (2004) 4379-4393.

\bibitem{egr} H. Egrifes and R. Sever, {\it Bound-State solutions of the Klein-Gordon equation for the generalized PT  -Symmetric Hulth\'en Potential}, Int. J. Theoret. Phys. 46 (2007)  935-950.


\bibitem{Gang} Gang Chen, Zi-Dong Chen and Zhi-Mei Lou, {\it Exact bound state solutions of the $s$-wave Klein–Gordon equation with the generalized Hulth\'en potential}, Phys. Lett. A 331 (2004) 374-377.

\bibitem{Benamira} F. Benamira, L. Guechi and A. Zouache, {\it Comment on ``Exact bound state solutions of the s-wave Klein-Gordon equation with the generalized Hulth\'en potential''}, Phys. Lett. A (2007), doi:10.1016/j.physleta.2007.05.089 . 

\bibitem{chen} Chang-Yuan Chen, Dong-Sheng Sun and Fa-Lin Lu, {\it Approximate analytical solutions of Klein–Gordon equation with Hulth\'en potentials for nonzero angular momentum}, Phys. Lett. A, doi:10.1016/j.physleta.2007.05.079.

\bibitem{Qiang} Wen-Chao Qiang, Run-Suo Zhou and Yang Gao, {\it Any $\ell$-state solutions of the Klein-Gordon equation with the generalized Hulth\'en potential}, Physics Letters A (2007), doi: 10.1016/j.physleta.2007.04.109.

\bibitem{greiner}W. Greiner, {\it Relativistic Quantum Mechanics. Wave Equations\/}, 3rd ed. (Springer, Berlin 2000).

\bibitem{alhaidari} A. D. Alhaidari, H Bahlouli and A. Al-Hasan, {\it Dirac and Klein-Gordon equations with equal scalar and vector potentials\/},  Phys. Lett. A 349 (2006) 87.
\bibitem{ikhdair} S. M. Ikhdair and R. Sever, {\it Approximate Eigenvalue and Eigenfunction Solutions for the Generalized Hulth\'en Potential with any Angular Momentum}, J. Math. Chem. (2006) DOI: 10.1007/s10910-006-9115-8. 

\bibitem{aktas} M. Aktas and R. Sever, {\it Exact supersymmetric solution of Schr\"odinger equation for central confining potentials by using the Nikiforov-Uvarov method} J. Mol. Struct. (Theochem) 710 (2004) 219-224.

\bibitem{filho} E. D. Filho and R. M. Ricotta, {\it Supersymmetry, Variational Method and Hulth\'en Potential} Mod. Phys. lett. A 10 (1995) 1613-1618.

\bibitem{gonul} B. G\"on\"ul, O. \"Ozer, Y. Cancelik, and M. Kocak, {\it Hamiltonian hierarchy and the Hulth\'en potential}, Phys. Lett. A 275 (2000) 238-243.

\bibitem{qian} S. W. Qian,  B. W. Huang and Z. Y. Gu, {\it Supersymmetry and shape invariance of the effective screened potential} New J. Phys. 4 (2002) 13.1-13.6.

\bibitem{gonu} B. G\"on\"ul, {\it Exact Treatment of $\ell\neq 0$ States,} Chin. Phys.  Lett.  21 (2003) 1685-1688.

\bibitem{Boztosun}  O Bayrak and I Boztosun, {\it Bound state solutions of the Hulth\'en potential by using the asymptotic iteration method}, Phys. Scr. 76 (2007) 92-96.

\bibitem{bayrak} O. Bayrak, G. Kocak and I. Boztosun, {\it Any $l$-state solutions of the Hulth\'en potential by the asymptotic iteration method,} J. Phys. A: Math. Gen. 39 (2006)  11521-11529.

\bibitem{Hakan} H. Ciftci, R. L. Hall, and N. Saad, {\it Construction of exact solutions to eigenvalue problems by the asymptotic iteration method\/}, J. Phys. A: Gen. Math. 38 (2005) 1147-1155.

\bibitem{andr} Larry C. Andrews, {\it Special functions of mathematics for engineers}, 2$^{th}$ edition, SPIE Press, Oxford Science Publication (1998), Chapter 11.

\bibitem{andrew} Larry C. Andrews, {\it Special functions of mathematics for engineers}, 2$^{th}$ edition, SPIE Press, Oxford Science Publication (1998), p. 378.

\bibitem{temme} Nico M. Temme, {\it Special functions: An introduction to the classical functions of mathematical physics}, John Wiley \& Sons Inc. New York (1996), Section 11.3.

\end{thebibliography}
\end{document}